\documentstyle[12pt]{article}
\begin{document}
\newlength{\lno}\lno1.5cm
\newlength{\len}\len=\textwidth\addtolength{\len}{-\lno}

\baselineskip6mm
\newcommand{\be}{\begin{equation}\label}\newcommand{\ee}{\end{equation}}
\newcommand{\bea}{\begin{eqnarray*}}\newcommand{\eea}{\end{eqnarray*}}
\def\eqa#1#2{\\\parbox{\len}{\bea#2\eea}\hfill\parbox{\lno}{\be{#1}\ee}}
\newcommand{\ban}{\begin{eqnarray}\label}\newcommand{\ean}{\end{eqnarray}}
\def\nn{\nonumber}
\newcommand{\bp}{\begin{picture}}\newcommand{\ep}{\end{picture}}
\newcommand{\ba}{\begin{array}}\newcommand{\ea}{\end{array}}
\newcommand{\s}{\scriptstyle}
\newcommand{\x}{{\underline x}}
\newcommand{\ua}{{\underline\alpha}}
\newcommand{\ub}{{\underline\beta}}
\newcommand{\uu}{{\underline u}}
\renewcommand{\theequation}{\mbox{\arabic{section}.\arabic{equation}}}
\newcommand{\la}{\langle\,}\newcommand{\ra}{\,\rangle}
\newcommand{\mi}{\,|\,}

\newtheorem{theo}{Theorem}[section]
\newtheorem{cond}[theo]{Conditions}
\newtheorem{defi}[theo]{Definition}
\newtheorem{exam}[theo]{Example}\newtheorem{lemm}[theo]{Lemma}
\newtheorem{rema}[theo]{Remark}
\newtheorem{prop}[theo]{Proposition}
\newtheorem{coro}[theo]{Corollary}
\newtheorem{beth}[theo]{BETHE ANSATZ}

\def\V{{V^{1\dots n}}}
\def\f{{f^{1\dots n}}}
\def\O{{\Omega^{1\dots n}}}
\def\VV{{{V^{(1)}}^{1\dots m}}}
\def\ff{{{f^{(1)}}^{1\dots m}}}

\def\T#1{{T_{1\dots n,#1}}}
\def\Tq#1{{T_{1\dots n,#1}^Q}}

\def\A{{A_{1\dots n}}}
\def\B{{B_{1\dots n}}}
\def\C{{C_{1\dots n}}}
\def\D{{D_{1\dots n}}}
\def\einsop{{\bf 1}}

\title{$SU(N)$ Matrix Difference Equations\\ and a Nested Bethe Ansatz}
\author{\\
H. Babujian$^{1,2,3}$, M. Karowski$^4$ and A. Zapletal$^{5,6}$
\\{\small\it Institut f\"ur Theoretische Physik}\\
{\small\it Freie Universit\"at Berlin, Arnimallee 14, 14195 Berlin, Germany} 
}
\date{\small\today}

\maketitle
\footnotetext[1]{Permanent address: Yerevan Physics Institute,
Alikhanian Brothers 2, Yerevan, 375036 Armenia.}
\footnotetext[2]{Partially supported by the grant 211-5291 YPI of the German
Bundesministerium f\"ur Forschung und Technologie and
through the VW project 'Cooperation with scientists from CIS'.}
\footnotetext[3]{e-mail: babujian@lx2.yerphi.am}
\footnotetext[4]{e-mail: karowski@physik.fu-berlin.de}
\footnotetext[5]{Supported by DFG, Sonderforschungsbereich 288
'Differentialgeometrie und Quantenphysik'}
\footnotetext[6]{e-mail: zapletal@physik.fu-berlin.de} 


\begin{abstract}
A system of $SU(N)$-matrix difference equations is solved by means of a
nested version of a generalized Bethe Ansatz, also called "off shell" Bethe
Ansatz \cite{B}. The highest weight property of the solutions is proved.
\end{abstract}

\section{Introduction}\label{s1}
Difference equations play a role in various contexts of mathematical
physics (see e.g. \cite{FR} and references therein). We are interested in
the application to
the form factor program in the exact integrable 1+1 dimensional field theory,
which was formulated in 1978 by one of the authors
(M.K.) and Weisz \cite{KW}. Form factors are matrix elements of local
operators
${\cal O}(x)$
$$
F(i\pi-\theta)=\la p'\mi{\cal O}(0)\mi p\ra
$$
where $p'p=m^2\cosh\theta$. Difference equations for these
functions are obtained by Watson's equations \cite{Wa}
\be{1.2}
F(\theta)=S(\theta)\,F(-\theta)~,~~~
F(i\pi-\theta)=F(i\pi+\theta)
\ee
where $S$ is the $S$-matrix.
For several models these equations have been solved in \cite{KW} and many
later publications (see e.g. \cite{Sm,ff} and references therein).
Generalized form factors are matrix elements for many particle states.
For generalized form factors Watson's equations lead typically to matrix
difference equations, which can be solved by a generalized Bethe Ansatz, 
also called "off-shell Bethe Ansatz". The conventional Bethe Ansatz
introduced by Bethe \cite{Be} is used to solve eigenvalue problems.
The algebraic formulation, which is also used in this article, was worked out
by Faddeev and coworkers (see e.g. \cite{TF}).
The "off-shell Bethe Ansatz" has been
introduced by one of the authors (H.B.) to solve the Knizhnik-Zamolodchikov
equations which are differential equations.
In \cite{Re} a variant of this technique has been formulated
to solve matrix difference equations of the form
$$
f(x_1,\dots,x_i+2,\dots,x_n)=Q(x_1,\dots,x_n,;i)\,f(x_1\dots,x_i,\dots,x_n)
~,~~(i=1,\dots,n)
$$
where $f(\x)$ is a vector valued function and the $Q(\x,i)$ are matrix
valued functions to be specified later.
For higher rank internal symmetry
groups the nested version of this Bethe Ansatz has to be applied.
The nested Bethe Ansatz to solve eigenvalue problems was introduced by
Yang \cite{Y} and further developed by Sutherland \cite{Su} (see also 
\cite{KR} for the algebraic formulation). The very interesting generalization
of this technique, which is applicable to difference
equations, is developed in this article for the $SU(N)$ - symmetry group.
This generalization demonstrates the power of the Bethe Ansatz even more
beautifully than the conventional applications.
Other methods to solve such matrix
difference equations have been discussed in \cite{Sm,Ma,Na,TV}.
In part II we will solve the $U(N)$ case, which will be used in \cite{BFK}
to solve the form factor problem for the $SU(N)$ chiral Gross-Neveu model.

The article is organized as follows.
In Section \ref{s2} we recall some well known results concerning the
$SU(N)$ R-matrix, the monodromy matrix and some commutation rules.
In Section \ref{s3} we introduce the nested generalized Bethe Ansatz to
solve a system of $SU(N)$ difference equations and present the solutions
in terms of "Jackson-type Integrals".
The proof of the main theorem avoids the decomposition of the monodromy
matrix, as used e.g. in \cite{Re}. Instead we introduce a new type of
monodromy matrix fulfilling a new type of Yang-Baxter relation and which is
adapted to the difference problem. In particular this yields an essential
simplification of the proof of the main theorem.
In Section \ref{s4} we prove the highest weight property of the solutions
and calculate the weights.
Section \ref{s5} contains some examples of solutions of the $SU(N)$
difference equations.

\section{The $SU(N)$ - R-matrix}\label{s2}
\setcounter{equation}{0}
Let $\V$ be the tensor product space
\be{2.1}\V=V_1\otimes\dots\otimes V_n\ee
where the vector spaces $V_i\cong{\bf C}^N,~(i=1,\dots,n)$ are considered as 
fundamental (vector) representation spaces of $SU(N)$. It is straightforward
to generalize the results of this paper to the case where the $V_i$ are
vector spaces for other representations.
We denote the canonical basis vectors by
\be{2.2}
\mi\alpha_1\dots\alpha_n\ra\in \V,~~(\alpha_i=1,\dots,N).
\ee
A vector $v^{1\dots n}\in \V$ is given in term of its
components by
\be{2.3}
v^{1\dots n}=\sum_\ua\mi\alpha_1\dots\alpha_n\ra\,
v^{\alpha_1,\dots,\alpha_n}.
\ee
A matrix acting in $\V$ by is denoted by
\be{2.4}\A~:~\V\to\V.\ee

The $SU(N)$ spectral parameter dependent R-matrix \cite{BKKW} acts on the
tensor product of two (fundamental) representation spaces of $SU(N)$. It
may be written and depicted as
\be{2.5}
R_{12}(x_1-x_2)=b(x_1-x_2)\,{\bf 1}_{12}+c(x_1-x_2)\,P_{12}~~=~~
\ba{c}
\unitlength2.7mm
\bp(5,4)
\put(1,0){\line(1,1){4}}
\put(5,0){\line(-1,1){4}}
\put(0,.5){$\s x_1$}
\put(5,.5){$\s x_2$}
\ep
\ea
~~~~:\,V^{12}\to V^{21},
\ee
where $P_{12}$ is the permutation operator.
Here and in the following we associate a variable (spectral parameter)
$x_i\in{\bf C}$ to each space $V_i$ which is graphically represented by a
line labeled by $x_i$ (or simply by $i$).
The components of the R-matrix are
\be{2.6}
R_{\alpha\beta}^{\delta\gamma}(x_1-x_2)=
\delta_{\alpha\gamma}\delta_{\beta\delta}\,b(x_1-x_2)+
\delta_{\alpha\delta}\delta_{\beta\gamma}\,c(x_1-x_2)~~=~
\ba{c}
\unitlength2mm
\bp(6,6)
\put(1,1){\line(1,1){4}}
\put(5,1){\line(-1,1){4}}
\put(.5,.1){$\s\alpha$}
\put(5,.1){$\s\beta$}
\put(5,5.3){$\s\gamma$}
\put(.5,5.3){$\s\delta$}
\put(0,2){$\s x_1$}
\put(4.8,2){$\s x_2$}
\ep
\ea~,
\ee
and the functions
\be{2.7}
b(x)=\frac x{x-2/N},~~c(x)=\frac{-2/N}{x-2/N}
\ee
are obtained as the rational solution of the Yang-Baxter equation
$$R_{12}(x_1-x_2)\,R_{13}(x_1-x_3)\,R_{23}(x_2-x_3)
=R_{23}(x_2-x_3)\,R_{13}(x_1-x_3)\,R_{12}(x_1-x_2),
$$
where we have employed the usual notation \cite{Y}. This relation is
depicted as
$$
\ba{c}
\unitlength4mm
\bp(9,4)
\put(0,1){\line(1,1){3}}
\put(0,3){\line(1,-1){3}}
\put(2,0){\line(0,1){4}}
\put(4.3,2){$=$}
\put(6,0){\line(1,1){3}}
\put(6,4){\line(1,-1){3}}
\put(7,0){\line(0,1){4}}
\put(.2,.5){$\s 1$}
\put(1.3,0){$\s 2$}
\put(3,.2){$\s 3$}
\put(5.5,.2){$\s 1$}
\put(7.2,0){$\s 2$}
\put(8.4,.4){$\s 3$}
\ep~~.
\ea
$$
The "unitarity" of the R-matrix reads and may depicted as
$$R_{21}(x_2-x_1)\,R_{12}(x_1-x_2)=1~:~~~~~
\ba{c}
\unitlength3mm
\bp(9,4)
\put(1,0){\line(1,1){2}}
\put(3,0){\line(-1,1){2}}
\put(1,2){\line(1,1){2}}
\put(3,2){\line(-1,1){2}}
\put(7,0){\line(0,1){4}}
\put(9,0){\line(0,1){4}}
\put(4.5,1.7){$=$}
\put(.2,0){$\s 1$}
\put(3.2,0){$\s 2$}
\put(6.2,0){$\s 1$}
\put(8.2,0){$\s 2$}
\ep
\ea~.
$$
As usual we define the monodromy matrix (with $\x=x_1,\dots,x_n$)
\be{2.8}
\T{0}(\x,x_0)=R_{10}(x_1-x_0)\,R_{20}(x_2-x_0)\dots
R_{n0}(x_n-x_0)=
\ba{c}
\unitlength2.5mm
\bp(10,4)
\put(0,2){\line(1,0){10}}
\put(2,0){\line(0,1){4}}
\put(4,0){\line(0,1){4}}
\put(8,0){\line(0,1){4}}
\put(1,0){$\s 1$}
\put(3,0){$\s 2$}
\put(7,0){$\s n$}
\put(9,1){$\s 0$}
\put(5,1){$\s\dots$}
\ep
\ea
\ee
as a matrix acting in the tensor product of the "quantum space"
$\V$ and the "auxiliary space" $V_0$ (all
$V_i\cong{\bf C}^N$). The Yang-Baxter algebra relations
\be{2.9}
\T{a}(\x,x_a)\,\T{b}(\x,x_b)\,R_{ab}(x_a-x_b)
=R_{ab}(x_a-x_b)\,\T{b}(\x,x_b)\,\T{a}(\x,x_a)
\ee
imply the basic algebraic properties of the sub-matrices w.r.t the
auxiliary space defined by
\be{2.10}
{T_{1\dots n}}^\alpha_\beta(\x,x)\equiv
\left(\matrix{\A(\x,x)&\B_\beta(\x,x)\cr
\C^\alpha(\x,x)&\D^\alpha_\beta(\x,x)}\right)~.
\ee
The indices $\alpha,\beta$ on the left hand side run from 1 to $N$ and on
the right hand side from 2 to $N$.
The commutation rules which we will need later are
\be{2.11}
\B_\alpha(\x,x')\,\B_\beta(\x,x)
=\B_{\beta'}(\x,x)\,\B_{\alpha'}(\x,x')\,
R^{\alpha'\beta'}_{\beta\alpha}(x-x'),
\ee
\ban{2.12}\nn
\A(\x,x')\,\B_\beta(\x,x)
&=&\frac1{b(x'-x)}\,\B_\beta(\x,x)\,\A(\x,x')\\
&&-\frac{c(x'-x)}{b(x'-x)}\,\B_\beta(\x,x')\,\A(\x,x)
\ean
and
\ban{2.13}\nn
\D_\gamma^{\gamma'}(\x,x')\,\B_\beta(\x,x)
&=&\frac1{b(x-x')}\,\B_{\beta'}(\x,x)\,\D^{\gamma'}_{\gamma''}(\x,x')\,
R^{\gamma''\beta'}_{\beta\gamma}(x-x')\\
&&-\frac{c(x-x')}{b(x-x')}\,
\B_\gamma(\x,x')\,\D^{\gamma'}_\beta(\x,x).
\ean

\section{The $SU(N)$ - difference equation}\label{s3}
\setcounter{equation}{0}
Let 
$$
\f(\x)=~~
\ba{c}
\unitlength4mm
\bp(4,3)
\put(2,1){\oval(4,2)}
\put(2,1){\makebox(0,0){$f$}}
\put(1,2){\line(0,1){1}}
\put(3,2){\line(0,1){1}}
\put(0,2.5){$\s x_1$}
\put(3.2,2.5){$\s x_n$}
\put(1.4,2.5){$\dots$}
\ep
\ea
~~\in \V
$$
be a vector valued function of $\x=x_1,\dots,x_n$ with values in
$\V$. The components of this vector are denoted by
$$f^{\alpha_1\dots\alpha_n}(\x)~,~~(1\le\alpha_i\le N).$$
\begin{cond}\label{cond}
The following symmetry and periodicity conditions of the vector
valued function $\f(\x)$ are supposed to be valid:
\begin{itemize}
\item[\rm(i)] The symmetry property under the exchange of two neighboring
spaces $V_i$ and $V_j$ and the variables $x_i$ and $x_j$, at the same time,
is given by
\be{3.1}
f^{\dots ji\dots}(\dots,x_j,x_i,\dots)=
R_{ij}(x_i-x_j)\,f^{\dots ij\dots}(\dots,x_i,x_j,\dots).
\ee
\item[\rm(ii)]
The {\bf system of matrix difference equations} holds
\be{3.2}
\fbox{\rule[-3mm]{0cm}{8mm} $
\f(\dots,x_i+2,\dots)=Q_{1\dots n}(\x;i)\,\f(\dots,x_i,\dots)
~,~~(i=1,\dots,n)$ }
\ee
where the matrices $Q_{1\dots n}(\x;i)\in End(\V)$ are defined by
\be{3.3}
Q_{1\dots n}(\x;i)=R_{i+1i}(x_{i+1}-x_i')\dots R_{ni}(x_n-x_i')\,
R_{1i}(x_1-x_i)\dots R_{i-1i}(x_{i-1}-x_i)
\ee
with $x'_i=x_i+2$.
\end{itemize}
\end{cond}
The Yang-Baxter equations for the R-matrix guarantee that these conditions
are compatible. 
The shift of $2$ in eq.~(\ref{3.2}) could be replaced by
an arbitrary $\kappa$. For the application to the form factor
problem, however, it is fixed to $2$ because of crossing symmetry.
Conditions \ref{cond} (i) and (ii) may be depicted as
\bea
{\rm(i)}&&~~~~~~
\ba{c}
\unitlength3mm
\bp(8,4)
\put(4,1){\oval(8,2)}
\put(4,1){\makebox(0,0){$f$}}
\put(1,2){\line(0,1){2}}
\put(3,2){\line(0,1){2}}
\put(5,2){\line(0,1){2}}
\put(7,2){\line(0,1){2}}
\put(1.5,2.5){$\s\dots$}
\put(5.5,2.5){$\s\dots$}
\ep
\ea
~~=~~
\ba{c}
\unitlength3mm
\bp(8,4)
\put(4,1){\oval(8,2)}
\put(4,1){\makebox(0,0){$f$}}
\put(1,2){\line(0,1){2}}
\put(3,2){\line(1,1){2}}
\put(5,2){\line(-1,1){2}}
\put(7,2){\line(0,1){2}}
\put(1.5,2.5){$\s\dots$}
\put(5.5,2.5){$\s\dots$}
\ep~,
\ea\\
{\rm(ii)}&&~~~~~~
\ba{c}
\unitlength3mm
\bp(8,5)
\put(4,2){\oval(7,2)}
\put(4,2){\makebox(0,0){$f$}}
\put(2,3){\line(0,1){2}}
\put(4,3){\line(0,1){2}}
\put(6,3){\line(0,1){2}}
\put(2.5,3.5){$\s\dots$}
\put(4.5,3.5){$\s\dots$}
\ep
\ea
~~=~~
\ba{c}
\unitlength3mm
\bp(8,5)
\put(4,2){\oval(7,2)}
\put(4,2){\makebox(0,0){$f$}}
\put(2,3){\line(0,1){2}}
\put(6,3){\line(0,1){2}}
\put(2.5,3.5){$\s\dots$}
\put(4.5,3.5){$\s\dots$}
\put(2,3){\oval(4,2)[t]}
\put(4,3){\oval(8,6)[b]}
\put(6,3){\oval(4,2)[tr]}
\put(6,5){\oval(4,2)[bl]}
\ep
\ea
\eea
with the graphical rule that a line changing the
"time direction" changes the spectral parameters $x\to x\pm 1$ as follows
$$
\ba{c}
\unitlength3mm
\bp(12,2)
\put(2,0){\oval(2,4)[t]}
\put(9,2){\oval(2,4)[b]}
\put(0,0){$\s x$}
\put(3.3,0){$\s x-1$}
\put(7,1){$\s x$}
\put(10.3,1){$\s x+1$}
\ep~.
\ea
$$
The $Q_{1\dots n}(\x;i)$ fulfill the commutation rules
\eqa{3.4}{
Q_{1\dots n}(\dots x_i\dots x_j+2\dots;i)\,
Q_{1\dots n}(\dots x_i\dots x_j\dots;j)~~~~~~~~~~~~\\
=Q_{1\dots n}(\dots x_i+2\dots x_j\dots;j)\,
Q_{1\dots n}(\dots x_i\dots x_j\dots;i).
}
The following Proposition is obvious
\begin{prop}\label{p3.1}
Let the vector valued function $\f(\x)\in\V$ fulfill Condition \ref{cond}
(i), then Conditions \ref{cond} (ii) for all $i=1,\dots,n$ are equivalent to
each other and also equivalent to the following periodicity property
under cyclic permutation of the spaces and the variables
\be{3.5}
f^{12\dots n}(x_1,x_2,\dots,x_n+2)=f^{n1\dots n-1}(x_n,x_1,\dots,x_{n-1}).
\ee
\end{prop}
\begin{rema}
The equations (\ref{3.1},\ref{3.5}) imply  Watson's (\ref{1.2}) equations
for the form factors \cite{BFK}.
\end{rema}

For later convenience we write the matrices 
\be{3.6}
Q_{1\dots n}(\x;i)={\rm tr}_0\,\Tq{0}(\x;i)
\ee
as the trace of of a new type of monodromy matrices where 
to the horizontal line two different spectral parameters are associated,
namely one to the right hand side and the other one to the left hand side.
However, both are related to a spectral parameter of one of the vertical
lines. This new monodromy matrix is given by the following
\begin{defi} 
For $i=1,\dots,n$
\be{3.7}
\Tq{0}(\x;i)=R_{10}(x_1-x_i)\dots R_{i-10}(x_{i-1}-x_i)\,P_{i0}\,
R_{i+10}(x_{i+1}-x'_i)\dots R_{n0}(x_n-x'_i)
\ee
$$~~=~~
\ba{c}
\unitlength10mm
\bp(6,1)
\put(1,0){\line(0,1){1}}
\put(2.5,0){\line(0,1){1}}
\put(3.5,0){\line(0,1){1}}
\put(5,0){\line(0,1){1}}
\put(0,0){\oval(6,1)[tr]}
\put(6,1){\oval(6,1)[bl]}
\put(.7,0){$\s x_1$}
\put(2.7,0){$\s x_i$}
\put(2.7,.9){$\s x'_i$}
\put(4.62,0){$\s x_n$}
\put(.5,.6){$\s x_i$}
\put(5.5,.65){$\s x'_i$}
\put(1.5,.7){$\dots$}
\put(4,.7){$\dots$}
\ep
\ea
$$
with $x'_i=x_i+2$.
\end{defi} 
Note that for $i=n$ one has simply $\Tq{0}(\x;n)
=\T{0}(\x,x_n)$ since $R(0)$ is the permutation operator $P$. 

The new type of monodromy matrix fulfills a new type of Yang-Baxter relation.
Instead of eq.~(\ref{2.9}) we have for $i=1,\dots,n$
\be{3.8}
\Tq{a}(\x;i)\,\T{b}(\x,u)\,R_{ab}(x'_i-u)
=R_{ab}(x_i-u)\,\T{b}(\x',u)\,\Tq{a}(\x;i)
\ee
with $\x'=x_1,\dots,x'_i,\dots,x_n$ and $x'_i=x_i+2$.
This relation follows from the Yang-Baxter equation for the R-matrix and
the obvious relation for the permutations operator $P$
$$
P_{ia}\,R_{ib}(x_i-u)\,R_{ab}(x'_i-u)=R_{ab}(x_i-u)\,R_{ib}(x'_i-u)\,P_{ia}.
$$
Correspondingly to eq.~(\ref{2.10}) we introduce (suppressing the indices
$1\dots n$)
\be{3.9}
{T^Q}^\alpha_\beta(\x;i)\equiv
\left(\matrix{A^Q(\x;i)&{B^Q}_\beta(\x;i)\cr
{C^Q}^\alpha(\x;i)&{D^Q}^\alpha_\beta(\x;i)}\right)~.
\ee
with the commutation rules with respect to the usual $A,B,C,D$
\be{3.10}
A^Q(\x;i)\,B_b(\x,u)=\frac1{b(x_i'-u)}\,B_b(\x',u)\,A^Q(\x;i)
-\frac{c(x_i'-u)}{b(x_i'-u)}\,{B^Q}_b(\x;i)\,A(\x,u),
\ee
\ban{3.11}\nn
{D^Q}_a(\x;i)\,B_b(\x,u)
&=&\frac1{b(u-x_i')}\,B_b(\x',u)\,{D^Q}_a(\x;i)\,R_{ba}(u-x'_i)\\
&&-\frac{c(u-x_i)}{b(u-x_i)}\,{B^Q}_b(\x;i)\,D_a(\x,u)\,P_{ab}.
\ean

The system of difference equations (\ref{3.2}) can be solved by means of a
generalized ("off-shell") nested Bethe Ansatz.
The first level is given by the
\begin{beth}\label{be}
\be{3.12}
\f(\x)=\sum_\uu~\B_{\beta_m}(\x,u_m)\dots\B_{\beta_1}(\x,u_1)\,
\O~g^{\beta_1\dots\beta_m}(\x,\uu)
\ee
$$
\ba{c}
\unitlength4mm
\bp(4,3)
\put(2,1){\oval(4,2)}
\put(2,1){\makebox(0,0){$f$}}
\put(1,2){\line(0,1){1}}
\put(3,2){\line(0,1){1}}
\put(0,2.5){$\s x_1$}
\put(3.2,2.5){$\s x_n$}
\put(1.5,2.5){$\s\dots$}
\ep
\ea
~~=\sum_\uu~~\ba{c}
\unitlength4mm
\bp(10,6)
\put(8,1){\oval(4,2)}
\put(8,1){\makebox(0,0){$g$}}
\put(1,2){\oval(12,2)[tr]}
\put(1,2){\oval(16,6)[tr]}
\put(2,2){\line(0,1){4}}
\put(3,2.5){$\s \dots$}
\put(3,5.5){$\s \dots$}
\put(7.5,2.5){$\s \dots$}
\put(1.2,3.5){$\s \vdots$}
\put(5,2){\line(0,1){4}}
\put(1,5.5){$\s x_1$}
\put(5.2,5.5){$\s x_n$}
\put(6,3.2){$\s u_1$}
\put(9.2,3.2){$\s u_m$}
\put(1.8,1.2){$\s 1$}
\put(4.8,1.2){$\s 1$}
\put(.3,2.8){$\s 1$}
\put(.3,4.8){$\s 1$}
\ep
\ea
$$
where summation over $\beta_1,\dots,\beta_m$ is assumed and $\O\in\V$
is the reference state defined by $\C^\beta\,\O=0$ for $1<\beta\le N$.
The summation over $\uu$ is specified by
\be{3.13}
\uu=(u_1,\dots,u_m)=(\tilde u_1-2l_1,\dots,\tilde u_m-2l_m)
~,~~~l_i\in{\bf Z},
\ee
where the $\tilde u_i$ are arbitrary constants.
\end{beth}
The reference state is
\be{3.14}\O=\mi 1\dots 1\ra,\ee
a basis vector with components
$\Omega^{\alpha_1\dots\alpha_n}=\prod_{i=1}^n\delta_{\alpha_i1}$.
It is an eigenstate of $\A$ and $\D$
\be{3.15}
\A(\x,u)\O=\O~,~~
\D^\alpha_\beta(\x,u)\O=\O\,\delta_{\alpha\beta}\prod_{i=1}^nb(x_i-u).
\ee
The sums (\ref{3.12}) are also called "Jackson-type Integrals" (see e.g.
\cite{Re} and references therein).
Note that the summations over $\beta_i$ run only over 
$1<\beta_i\le N$. Therefore the $g^{\beta_1\dots\beta_m}$ are the
components
of a vector $g^{1\dots m}$ in the tensor product of smaller spaces
${V^{(1)}}^{1\dots m}=V^{(1)}_1\otimes\dots\otimes V^{(1)}_m$ with
$V^{(1)}_{i}\cong{\bf C}^{N-1}$.
\begin{defi}\label{d3.2}
Let the vector valued function $\ff(\uu)\in\VV$ be given by
\be{3.16}
g^{1\dots m}(\x,\uu)=\prod_{i=1}^n\prod_{j=1}^m\psi(x_i-u_j)\,
\prod_{1\le i<j\le m}\tau(u_i-u_j)\,\ff(\uu)
\ee
where the functions $\psi(x)$ and $\tau(x)$ fulfill the functional equations
\be{3.17}
b(x)\,\psi(x)=\psi(x-2)~,~~~
\frac{\tau(x)}{b(x)}=\frac{\tau(x-2)}{b(2-x)}.
\ee
\end{defi}
Using the definition of $b(x)$ (\ref{2.7}) we get the solutions of
eqs.~(\ref{3.17})
\be{3.18}
\psi(x)=\frac{\Gamma(1-\frac1N+\frac x2)}{\Gamma(1+\frac x2)}
,~~~\tau(x)=\frac{x\,\Gamma(\frac1N+\frac x2)}
{\Gamma(1-\frac1N+\frac x2)}
\ee
where the general solutions are obtained by multiplication with
arbitrary periodic functions with period 2.
Just as $g^{1\dots m}(\x,\uu)$ also the vector valued function
$\ff(\uu)$ is an element of the tensor product of the smaller
spaces $V_i\cong{\bf C}^N$
$$\ff(\uu)\in\VV.$$
We say $\ff(\uu)$ fulfills Conditions \ref{cond} (i)$^{(1)}$
and  (ii)$^{(1)}$ if eqs.~(\ref{3.1}) and (\ref{3.2}) hold in this space.
We are now in a position to formulate the main theorem of this paper.
\begin{theo}\label{t3.1}
Let the vector valued function $\f(\x)$ be given by the Bethe Ansatz
\ref{be} and let $g^{1\dots m}(\x,\uu)$ be of the form of Definition
\ref{d3.2}.
If in addition the vector valued function $\ff(\uu)\in\VV$ fulfills the
Conditions \ref{cond} {\rm(i)$^{(1)}$ and (ii)$^{(1)}$}, then also
$\f(\x)\in\V$ fulfills the Conditions \ref{cond} {\rm(i) and (ii)}, i.e.
$\f(\x)$ is a solution of the set of difference equations (\ref{3.2}).
\end{theo}
\begin{rema}
For $SU(2)$ (see e.g.\cite{Re}) the problem is already solved by \ref{3.16}
since then $f^{(1)}$ is a constant.
\end{rema}
{\bf Proof:} Condition \ref{cond} (i) follows directly from the definition
and the normalization of the R-matrix (\ref{2.5})
$$R_{ij}(x_i-x_j)\,\Omega^{\dots ij\dots}=\Omega^{\dots ij\dots},$$
the symmetry of $g^{1\dots m}(\x,\uu)$ given by eq.~(\ref{3.16}) under
the exchange of $x_1,\dots,x_n$ and
$$
B_{\dots ji\dots,\beta}(\dots x_j,x_i\dots,u)\,R_{ij}(x_i-x_j)=
R_{ij}(x_i-x_j)\,B_{\dots ij\dots,\beta}(\dots x_i,x_j\dots,u)
$$
which is a consequence of the Yang-Baxter relations for the R-matrix.

Because of Proposition \ref{p3.1} it is sufficient to prove Condition
\ref{cond} (ii) only for $i=n$
$$
Q(\x;n)\,f(\x)=
{\rm tr}_a\,T^Q_a(\x;n)\,f(\x)=f(\x')~,~~(\x'=x_1,\dots,x'_n=x_n+2).
$$
where the indices $1\dots n$ have been suppressed.
For the first step we apply a technique quite analogous to that used for the
conventional algebraic Bethe Ansatz which solves eigenvalue problems.
We apply the trace of $T^Q_a(\x;n)$ to the vector
$f(\x)$ as given by eq.~(\ref{3.12}) and push $A^Q(\x;n)$ and $D^Q_a(\x;n)$
through all the $B$'s using the commutation rules (\ref{3.10}) and
(\ref{3.11}). Again with $\x'=x_1,\dots,x'_n=x_n+2$ we obtain
\bea
A^Q(\x;n)\,B_{b_m}(\x,u_m)\dots B_{b_1}(\x,u_1)
=B_{b_m}(\x',u_m)\dots B_{b_1}(\x',u_1)
\prod_{j=1}^m\frac1{b(x'_n-u_j)}~~~~~~\\\times A^Q(\x;n)+{\rm uw}_A
\eea
\bea
D^Q_a(\x;n)\,B_{b_m}(\x,u_m)\dots B_{b_1}(\x,u_1)
=B_{b_m}(\x',u_m)\dots B_{b_1}(\x',u_1)
\prod_{j=1}^m\frac1{b(u_j-x_n)}~~~~~~\\
\times D^Q_a(\x;n)\,R_{b_1a}(u_1-x'_n)\dots R_{b_ma}(u_m-x'_n)+{\rm uw}_{D_a}
\eea
The "wanted terms" written explicitly originate from the first term in the
commutations rules (\ref{3.10}) and (\ref{3.11}); all other contributions
yield the "unwanted terms".
If we insert these equations into the representation (\ref{3.12}) of
$f(\x)$ we find that the wanted contribution from $A^Q$ already gives
the desired result. The wanted contribution from $D^Q$ applied to $\Omega$
gives zero. The unwanted contributions can be written as as difference
which vanishes after summation over the $u$'s. These
three facts can be seen as follows. We have
$$A^Q(\x;n)\,\Omega=\Omega~, ~~~D^Q_a(\x;n)\,\Omega=0$$
which follow from eq.~(\ref{3.15}) since $T^Q(\x;n)=T(\x,x_n)$ and $b(0)=0$.
The defining relation of $\psi(x)$
(\ref{3.17}) implies that the wanted term from $A$ yields $f(\x')$.
The commutation relations (\ref{3.10}), (\ref{3.11}), (\ref{2.12}) and
(\ref{2.13}) imply that
the unwanted terms are proportional to a product of $B$-operators, where
exactly one $B_{b_j}(\x,u_j)$ is replaced by $B^Q_{b_j}(\x;n)$.
Because of the commutation relations of the $B$'s (\ref{2.11}) and the
symmetry property given by Condition \ref{cond} (i)$^{(1)}$ of
$g^{1\dots m}(\x,\uu)$ it is sufficient to
consider only the unwanted terms for $j=m$ denoted by uw$_A^m$ and uw$_D^m$.
They come from the second term in (\ref{3.10}) if $A^Q(\x;n)$ is commuted
with $B_{b_m}(\x,u_m)$ and then the resulting $A(\x,u_m)$ pushed through the
other $B$'s taking only the first terms in (\ref{2.12}) into account
and correspondingly for $D^Q_a(\x;u_m)$.
$$
{\rm uw}_A^m=-\frac{c(x'_n-u_m)}{b(x'_n-u_m)}\,B^Q_{b_m}(\x;m)\dots
B_{b_1}(\x,u_1)\prod_{j<m}\frac1{b(u_m-u_j)}\,A(\x,u_m)
$$
$$
{\rm uw}_{D_a}^m=-\frac{c(u_m-x_n)}{b(u_m-x_n)}\,B^Q_{b_m}(\x;m)\dots
B_{b_1}(\x,u_1)\prod_{j<m}\frac1{b(u_j-u_m)}\,D_a(\x,u_m)
\,T^{Q(1)}_{b_1\dots b_m,a}(\uu;m)
$$
where $T^{Q(1)}$ is the new type of monodromy matrix
\be{3.18a}
T^{Q(1)}_{b_1\dots b_m,a}(\uu;m)=
R_{b_1a}(u_1-u_m)\dots R_{b_{m-1}a}(u_{m-1}-u_m)\,P_{b_ma}
\ee
analogous to (\ref{3.6})
whose trace over the auxiliary space $V^{(1)}_a$ yields the shift operator
$Q^{(1)}(\uu;m)$.
With $D_a(\x,u_m)\Omega=\einsop_a\,\prod_{i=1}^n
b(x_i-u_m)\,\Omega$ (see (\ref{3.15})), by the assumption
$$
Q^{(1)}(\uu;m)\,f^{(1)}(\uu)=f^{(1)}(\uu')~,~~~
(\uu'=u_1,\dots,u'_m=u_m+2)
$$
and the defining relations (\ref{3.17}) of $\psi(x)$ and $\tau(x)$,
we obtain
$$
{\rm tr}_a\,{\rm uw}_{D_a}^m(\uu)\,\Omega\,g(\x,\uu)=
-{\rm uw}_A^m(\uu')\,\Omega\,g(\x,\uu')
$$
where $c(-x)/b(-x)=-c(x)/b(x)$ has been used.
Therefore the sum of all unwanted terms yield a difference analog of a total
differential which vanishes after summation over the $u$'s.

Iterating Theorem \ref{t3.1} we obtain the nested generalized Bethe Ansatz
with levels $k=1,\dots,N-1$.
The Ansatz of level $k$ reads
\ban{3.19}
\lefteqn{{f^{(k-1)}}^{1\dots n_{k-1}}\left(\x^{(k-1)}\right)
=\sum_{\x^{(k)}}\,B^{(k-1)}_{1\dots n_{k-1}\beta_{n_k}}
\left(\x^{(k-1)},x^{(k)}_{n_k}\right)\dots}\\
&&~~~\dots B^{(k-1)}_{1\dots n_{k-1}\beta_1}\left(\x^{(k-1)},x^{(k)}_1
\right)\,{\Omega^{(k-1)}}^{1\dots n_{k-1}}~
{g^{(k-1)}}^{\beta_1\dots\beta_{n_k}}\left(\x^{(k-1)},\x^{(k)}\right)
\nn
\ean
The functions $f^{(k)}$ and $g^{(k)}$ are vectors with
$$
{f^{(k)}}^{1\dots n_k},{g^{(k-1)}}^{1\dots n_k}\in {V^{(k)}}^{1\dots n_k}
=V^{(k)}_1\otimes\dots\otimes V^{(k)}_{n_k}
~,~~~~(V^{(k)}_{i}\cong{\bf C}^{N-k}).
$$
The basis vectors of these spaces are $\mi\alpha_1\dots\alpha_{n_k}\ra^{(k)}
\in{V^{(k)}}^{1\dots n_k}$ and $k<\alpha_i\le N$.

Analogously to Definition \ref{d3.2} we write
\be{3.20}
{g^{(k-1)}}^{1\dots n_k}(\x^{(k-1)},\x^{(k)})
=\prod_{i=1}^{n_{k-1}}\prod_{j=1}^{n_k}\psi(x_i^{(k-1)}-x_j^{(k)})\,
\prod_{1\le i<j\le n_k}\tau(x^{(k)}_i-x^{(k)}_j)\,
{f^{(k)}}^{1\dots n_k}(\x^{(k)})
\ee
where the functions $\psi(x)$ and $\tau(x)$ fulfill the functional equations
(\ref{3.17}) with the solutions (\ref{3.18}).
Then the start of the iteration is given by a $k_{max}\le N$ with
\be{3.21}
{f^{(k_{max}-1)}}^{1\dots n_{n_{k_{max}-1}}}=\mi k_{max}\dots k_{max}\ra~,~~~
{\rm and}~~n_k=0~~{\rm for}~k\ge k_{max}
\ee
which is the reference state of level $k_{max}-1$ and trivially fulfills
the Conditions \ref{cond}.
\begin{coro}\label{c3.2}
The system of $SU(N)$ matrix difference equations (\ref{3.2})
is solved by the nested Bethe Ansatz (\ref{3.19}) with
(\ref{3.20}), (\ref{3.21}) and $\f(\x)={f^{(0)}}^{1\dots n}(\x)$.
\end{coro}

\section{Weights of generalized $SU(N)$ Bethe vectors}\label{s4}
\setcounter{equation}{0}
In this section we analyze some group theoretical properties of generalized
Bethe states.
We calculate the weights of the states and show that they are highest weight
states. The first result does not depend on any restriction to the states.
On the other hand the second result is not only true for the conventional
Bethe Ansatz, which solves an eigenvalue problem and which is well known,
but also, as we will show, for the generalized one which solves a
difference equation (or a differential equation).

By asymptotic expansion of the $R$-matrix and the monodromy
matrix $T$ (cf. eqs.(\ref{2.5}) and(\ref{2.8})) we get for $u\to\infty$
\ban{4.1}
R_{ab}(u)&=&\einsop_{ab}-\frac2{Nu}\,P_{ab}+ O(u^{-2})\\
\T{a}(\x,u)&=&\einsop_{1\dots n,a}+\frac2{Nu}\,M_{1\dots n,a}
+ O(u^{-2}).
\ean
Explicitly we get from eq.~(\ref{2.8})
\be{4.3}
M_{1\dots n,a}=P_{1a}+\dots+P_{na}
\ee
where the $P$'s are the permutation operators.
The matrix elements of $M_{1\dots n,a}$ as a matrix in the auxiliary space
are the $su(N)$ Lie algebra generators.
In the following we will consider only operators acting in the fixed tensor
product space $V=\V$ of (\ref{2.1}); therefore we will omit the
indices $1\dots n$.
In terms of matrix elements in the auxiliary space $V_a$ the generators
act on the basis states as
\be{4.4}
M^{\alpha'}_\alpha\mi\alpha_1,\dots,\alpha_i,\dots,\alpha_n\ra=\sum_{i=1}^n
\delta_{\alpha'\alpha_i}\mi\alpha_1,\dots,\alpha,\dots,\alpha_n\ra.
\ee

The Yang-Baxter relations (\ref{2.9}) yield for $x_a\to\infty$
\be{4.5}[M_a+P_{ab},T_b(x_b)]=0\ee
and if additionally $x_b\to\infty$
\be{4.6}[M_a+P_{ab},M_b]=0\ee
or for the matrix elements
\ban{4.7}
[M^{\alpha'}_\alpha,T^{\beta'}_\beta(u)]&=&
\delta_{\alpha'\beta}\,T^{\beta'}_\alpha(u)
-\delta_{\alpha\beta'}\,T^{\alpha'}_\beta(u)\\~\label{4.8}
 [M^{\alpha'}_\alpha,M^{\beta'}_\beta]&=&
\delta_{\alpha'\beta}\,M^{\beta'}_\alpha
-\delta_{\alpha\beta'}\,M^{\alpha'}_\beta.
\ean
Equation (\ref{4.8}) represents the structure relations of the $su(N)$ Lie
algebra and (\ref{4.7}) the $SU(N)$-covariance of $T$.
In particular the transfer matrix is invariant
\be{4.9}
[M^{\alpha'}_\alpha,{\rm tr}\,T(u)]=0.
\ee

We now investigate the action of the lifting operators
$M^{\alpha'}_\alpha$ ($\alpha'>\alpha$) to generalized Bethe vectors.
\begin{lemm}\label{l4.1}
Let $F[g](\x)\in V$ be a Bethe Ansatz vector given in terms of
a vector $g(\x,\uu)\in V^{(1)}\cong{\bf C}^{(N-1)\otimes m}$ by
\be{4.10}
F[g](\x,\uu)=B_{\beta_m}(\x,u_m)\dots B_{\beta_1}(\x,u_1)\,\Omega~
g^\ub(\x,\uu)
\ee
with $\ub=\beta_1,\dots,\beta_m$. Then $M^{\alpha'}_\alpha\,F[g]$
is of the form
\be{4.11}
M^{\alpha'}_\alpha\,F[g]=\cases{
\sum_{j+1}^mB_{\beta_m}\dots\delta_{\alpha'\beta_m}\dots B_{\beta_1}\,
\Omega~G_j^\ub(\x,\uu)&\rm for $\alpha'>\alpha=1$\cr
F[{M^{(1)}}^{\alpha'}_\alpha\,g]&\rm for $\alpha'>\alpha>1$,}
\ee
where the ${M^{(1)}}^{\alpha'}_\alpha$ are the $su(N-1)$ generators
represented in $V^{(1)}$ (analogously to (\ref{4.3})) and 
\be{4.12}
G_m(\x,\uu)=\left(\frac1{\prod_{j=1}^mb(u_m-u_j)}-
\frac{\prod_{i=1}^nb(x_i-u_m)}{\prod_{j=1}^mb(u_j-u_m)}\,
Q^{(1)}(\uu;m)\right)g(\x,\uu).
\ee
The operator $Q^{(1)}(\uu;m)\in End(V^{(1)})$ is a next level Q-matrix given
by the trace
\be{4.13}\nn
Q^{(1)}(\uu;m)={\rm tr}_a\,T^{Q(1)}_{a}(\uu;m)
\ee
(see eq.~(\ref{3.18a})). The other $G_j$ are obtained by Yang-Baxter
relations.
\end{lemm}
{\bf Proof:}
First we consider the case $\alpha=1$. The commutation rule (\ref{4.7})
reads for $\beta'=1$ and $\alpha'\to\alpha$
$$[M^\alpha_1,B_\beta(u)]=\delta_{\alpha\beta}\,A(u)-D^\alpha_\beta(u).$$
We commute $M^\alpha_1$ through all the $B$'s of (\ref{4.10}) and use
$M^\alpha_1\,\Omega=0$ for $\alpha>1$ (cf.~(\ref{4.4})).
The $A$'s and $D$'s appearing are also commuted through all the $B$'s using
the commutation rules (\ref{2.12}) and (\ref{2.13}). In each summand exactly
one $B$-operator disappears. Therefore the result is of the form of
eq.~(\ref{4.11}). Contributions to $G_m$ arise when we commute $M^\alpha_1$
through $B_{\beta_m}(u_m)$ and then push the $A(u_m)$ and $D(u_m)$ through
the other $B$'s ($j<m$), only taking the first terms of (\ref{2.12}) and
(\ref{2.13}) into account. All other terms would contain a $B(u_m)$ and
would therefore contribute to one of the other $G_j~(j<m)$. Finally we
apply $A(u_m)$ and $D(u_m)$ to $\Omega$
$$A(u_m)\,\Omega=\Omega~,~~~
D^\alpha_\beta(u_m)\,\Omega=\delta_{\alpha\beta}\,
\prod_{i=1}^nb(x_i-u_m)\,\Omega
$$
and get eq.~(\ref{4.12}).

For $\alpha'>\alpha>1$ we again use the commutation rule (\ref{4.7})
$$[M^{\alpha'}_\alpha,B_\beta(u)]=\delta_{\alpha'\beta}\,B_\alpha(u)$$
and get
$$
M^{\alpha'}_\alpha\,B_\beta(u_m)\dots B_\beta(u_1)
=B_{\beta_m}(u_m)\dots B_{\beta_1}(u_1)\,M^{\alpha'}_\alpha\\
+B_{\beta'_m}(u_m)\dots B_{\beta'_1}(u_1)\,
{M^{(1)}}^{\ub',\alpha'}_{\ub,\alpha}
$$
with $M^{(1)}_{1\dots m,a}=P^{(1)}_{1a}+\dots+P^{(1)}_{ma}$ analogously to
(\ref{4.3}). Because of $M^{\alpha'}_\alpha\,\Omega=0$ for $\alpha'>1$
(cf.~(\ref{4.4})) we get eq.~(\ref{4.11})

The diagonal elements of $M$ are the weight operators
$W_\alpha=M^\alpha_\alpha$, they act on the basis vectors in $V$ as
\be{4.14}
W_\alpha\mi\alpha_1,\dots,\alpha_n\ra=
\sum_{i=1}^n\delta_{\alpha_i\alpha}\,\mi\alpha_1,\dots,\alpha_n\ra
\ee
which follows from ${P_i}^{\alpha'}_\alpha\mi\alpha_i\ra=
\delta_{\alpha\alpha_i}\,\mi\alpha'\ra$.
In particular we get for the Bethe Ansatz reference state (\ref{3.14})
\be{4.15}W_\alpha\,\Omega=\delta_{\alpha1}\,n\,\Omega.\ee
\begin{lemm}\label{l4.2}
Let $F[g]\in\V$ be as in Lemma \ref{l4.1}. Then
\be{4.16}
W_\alpha\,F[g]=\cases{
(n-m)\,F[g]&\rm for $\alpha=1$\cr
F[W^{(1)}_\alpha\,g]&\rm for $\alpha>1$,}
\ee
where the $W^{(1)}_\alpha$'s are the $su(N-1)$ weight operators acting in
$V^{(1)}$, i.e. the diagonal
elements of generator matrix ${M^{(1)}}^{\alpha'}_\alpha$
(analogously to (\ref{4.3})).
\end{lemm}
{\bf Proof:}
By means of the commutation relation
(\ref{4.7}) for $\alpha'=\alpha=\beta'=1,~\beta>1$
$$[W_1,B_\beta]=-B_\beta$$
we commute $W_1$ through all $m~B$'s of eq.~(\ref{4.10}) and with
(\ref{4.15}) we get the first equation.
For the second equation we again use (\ref{4.7}) now for
$\alpha'=\alpha>1,~\beta'=1,~\beta>1$
$$[W_\alpha,B_\beta]=\delta_{\alpha\beta}\,B_\beta.$$
Again commuting $W_\alpha$ through all the $B$'s of eq.~(\ref{4.10}) we get
with eq.~(\ref{4.15})
\bea
W_\alpha\,B_{\beta_m}\dots B_{\beta_1}\,\Omega~g^{\beta_1\dots\beta_m}
&=&B_{\beta_m}\dots B_{\beta_1}\,\left(W_\alpha+\sum_{i=1}^m
\delta_{\beta_i\alpha}\right)\Omega~g^{\beta_1\dots\beta_m}\\
&=&B_{\beta_m}\dots B_{\beta_1}\,\Omega~
\Big(W^{(1)}_\alpha\,g\Big)^{\beta_1\dots\beta_m}
\eea
which concludes the proof.
\begin{theo}
Let the vector valued function $f(\x)\in V$ be given by the BETHE ANSATZ
\ref{be} fulfilling the assumptions of Theorem \ref{t3.1}.
If in addition $f^{(1)}$ is a highest weight vector and an eigenvector
of the weight operators with
\be{4.20}
W^{(1)}_\alpha\,f^{(1)}=w^{(1)}_\alpha\,f^{(1)},
\ee
then also $f$ is a highest weight vector
\be{4.17}
M^{\alpha'}_\alpha\,f=0~,~~(\alpha'>\alpha)
\ee
and an eigenvector of the weight operators
\be{4.18}
W_\alpha\,f=w_\alpha\,f~,~~~
w_\alpha=\cases{
n-m &\rm for $\alpha=1$\cr
w^{(1)}_\alpha &\rm for $\alpha>1$}
\ee
with
\be{4.19}
w_\alpha\ge w_\beta~,~~~(1\le\alpha<\beta\le N).
\ee
\end{theo}
{\bf Proof:}
To prove the highest weight property we apply Lemma \ref{l4.1}. By assumption
$f^{(1)}$ fulfills the difference equation
$$f^{(1)}(u_1,\dots,u_m+2)=Q^{(1)}(\uu;m)\,f^{(1)}(u_1,\dots,u_m).$$
Together with eq.~(\ref{3.16}), (\ref{3.17}) and (\ref{4.12}) we obtain
after summation $\sum_{u_m}G_m(\x,\uu)=0$, if
$u_m=\tilde u_m-2l_m~(l_m\in{\bf Z})$. The same is true for the other
$G_i$ in eq.~(\ref{4.11}), since $g$ fulfills the symmetry
property of Condition \ref{cond} (i) and thereby $F[g](\x,\uu)$ of
eq.~(\ref{4.10}) is symmetric with respect to the $u_i$.
Therefore in eq.~(\ref{4.12}) we have
$M^{\alpha'}_\alpha\,f=0$ for $\alpha'>\alpha>1$ and for $\alpha'>\alpha=1$
by assumption on $f^{(1)}$. The  weights of $f$ follow from Lemma \ref{l4.2}
and also by assumption on $f^{(1)}$.
From the commutation rule (\ref{4.8}) and ${M^\beta_\alpha}^\dagger=
M_\beta^\alpha$ follows
$$
0\le M^\beta_\alpha\,M_\beta^\alpha
=M_\beta^\alpha\,M^\beta_\alpha+W_\alpha-W_\beta
$$
which implies (\ref{4.19}).

Since the states $f^{(k_{max}-1)}$ of eq.~(\ref{3.21}) are
highest weight states in $V^{(k_{max}-1)}$
with weight $w^{(k_{max}-1)}_{k_{max}}=n_{k_{max}-1}$ we have the
\begin{coro}
If $f(\x)$ is a solution of
the system of $SU(N)$ matrix difference equations (\ref{3.2})
$$
f(\dots,x_i+2,\dots)=Q(\x;i)\,f(\dots,x_i,\dots)~,~~(i=1,\dots,n)
$$
given by the generalized nested Bethe Ansatz of Corollary \ref{c3.2},
then $f$ is a highest weight vector with weights
\be{4.21}
w=(w_1,\dots,w_N)=(n-n_1,n_1-n_2,\dots,n_{N-2}-n_{N-1},n_{N-1}),
\ee
where $n_k$ is the number of $B^{(k-1)}$ operators in the Bethe Ansatz of
level $k,~(k=1,\dots,N-1)$.
Further non-highest weight solutions of (\ref{3.2}) are given by
\be{4.22}
f^{\alpha'}_\alpha=M^{\alpha'}_\alpha\,f~,~~(\alpha'<\alpha).
\ee
\end{coro}
The interpretation of eq.~(\ref{4.21}) is that each $B^{(k)}$-operator
reduces $w_k$ and lifts a $w_l~(l>k)$ by one.

\section{Examples}\label{s5}
\setcounter{equation}{0}
From a solution of the matrix difference equations (\ref{3.2}) one gets a
new solution by multiplication of a scalar function which is symmetric
with respect to all variables $x_i$ and periodic with period $2$. Therefore
the solutions of the following examples may be multiplied by such functions.

\begin{exam}
The simplest example is obtained for $k_{max}=1$ which means the trivial
solution of the difference equations
$$\f=\O$$
The weights of $\f$ are $w=(n,0,\dots,0)$.
\end{exam}
In the language of spin chains this case corresponds to the ferro-magnetic
ground state.

\begin{exam}\label{e5.2}
For the case $k_{max}=1$ and $n^{(1)}=1$ the solution reads
$$
\f(\x)=\sum_u~\B_{,\beta}(\x,u)\,\O~g^\beta(\x,u).
$$
with $u=\tilde u-2l~(l\in{\bf Z},~\tilde u$ an arbitrary constant) and
$$
g^\beta(\x,u)=\delta_{\beta2}\prod_{i=1}^n\psi(x_i-u).
$$
The weights of this vector $\f$ are $w=(n-1,1,0,\dots,0)$.
The action of the creation operator $\B_\beta(x,y;u)$
on the reference state is easily calculated with help of
eqs.~(\ref{2.6}) (\ref{2.8}) and (\ref{2.10}).
\end{exam}
As a particular case of this example we determine explicitly the solution
for the following
\begin{exam}\label{e5.3}
The action of the $B$-operator 
on the reference state for the case of $n=2$ of Example \ref{e5.2} yields
$$
B_{12,\beta}(x,y;u)\mi11\ra=
c(x-u)\,b(y-u)\,\mi\beta1\ra+c(y-u)\,\mi1\beta\ra.
$$
Therefore we obtain
$$
f^{12}(x,y)=\sum_u\,\psi(x-u)\,\psi(y-u)\{
c(x-u)\,b(y-u)\,\mi\beta1\ra+c(y-u)\,\mi1\beta\ra\}
$$
with $u=\tilde u-2l,~(l\in{\bf Z})$.
Using the expressions for the functions $b,~c,~\psi$ given by
eqs.~(\ref{2.7}) and (\ref{3.18}) we get up to a
constant
$$
f^{12}(x,y)=\Big(
\sin\pi\left(\s\frac{x-\tilde u}2-\frac1N\right)\,
\sin\pi\left(\s\frac{y-\tilde u}2-\frac1N\right)\,
\Gamma\left(\s\frac{y-x}2-\frac1N\right)\,
\Gamma\left(\s1+\frac{x-y}2-\frac1N\right)\Big)^{-1}
(\mi21\ra-\mi12\ra).
$$
This solution could also be obtained by means of the method used
in \cite{KW}, namely by diagonalization of the R-matrix. One obtains the
scalar difference equations
$$
f_-(x,y)=R_-(x-y)\,f_-(y,x)~,~~~f_-(x,y)=f_-(y,x+2)
$$
with the eigenvalue $R_-(x)=(x+2/N)(x-2/N)$ of the antisymmetric tensor
representation.
\end{exam}

\begin{exam}\label{e3.1}
Next we consider for $N>2$ the case of the
quantum space $V_{123}=V_1\otimes V_2\otimes V_3$ and the case
that the nested Bethe Ansatz has only two
levels with two creation operators in the first level and one in
the second level.
This means $k_{max}=3,~n=3,~n^{(1)}=2,~n^{(2)}=1$ and the weights
$w=(1,1,1,0,\dots,0)$.
The first level Bethe Ansatz is given by
$$
f^{123}(x,y,z)=\sum_{u,v}~B_{123,\beta}(x,y,z;v)\,B_{123,\alpha}(x,y,z;u)
\,\Omega^{123}~g^{\alpha\beta}(x,y,z;u,v).
$$
where the summation is specified by
$u=\tilde u-2k,\,v=\tilde u-2l,~(k,l\in{\bf Z})$.
By eq.~(\ref{3.16})  $g^{12}$ is related to the next level function
${f^{(1)}}^{12}$ by 
$$
g^{12}(x,y,z;u,v)=\prod_{x_i=x,y,z}\prod_{u_j=u,v}\psi(x_i-u_j)
\,\tau(u-v)\,{f^{(1)}}^{12}(u,v).
$$
The second level Bethe Ansatz reads
$$
{f^{(1)}}^{12}(u,v)=\sum_w\,B^{(1)}_{12\gamma}(u,v;w)\,
{\Omega^{(1)}}^{12}~{g^{(1)}}^{\gamma}(u,v;w)
$$
where $w=\tilde w-2m,~(m\in{\bf Z})$.
The second level reference state is
${\Omega^{(1)}}^{12}=\mi 22\ra^{(1)}\in {V^{(1)}}^{12}$.
Again according to eq.~(\ref{3.16})
$$
{g^{(1)}}^\gamma(u,v;w)=\psi(u-w)\,\psi(v-w)\,{f^{(2)}}^\gamma,
$$
with ${f^{(2)}}^\gamma=\delta_{\gamma3}.$
Similar as in Example \ref{e5.3} the action of the operators $B$ and
$B^{(1)}$ on their reference states may be calculated.
\end{exam}
For this example the two level nested Bethe Ansatz may be depicted as
$$
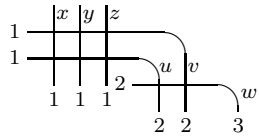
\ba{c}
\unitlength3,5mm
\bp(10,5)
\put(5,1){\oval(2,4)[tr]}
\put(6,1){\oval(2,6)[tr]}
\put(5,1){\oval(8,2)[tr]}
\put(1,3){\line(1,0){4}}
\put(1,4){\line(1,0){5}}
\put(2,2){\line(0,1){3}}
\put(3,2){\line(0,1){3}}
\put(4,2){\line(0,1){3}}
\put(2.1,4.5){$\s x$}
\put(3.1,4.5){$\s y$}
\put(4.1,4.5){$\s z$}
\put(6,2.5){$\s u$}
\put(7.1,2.5){$\s v$}
\put(9.1,1.5){$\s w$}
\put(1.8,1.2){$\s 1$}
\put(2.8,1.2){$\s 1$}
\put(3.8,1.2){$\s 1$}
\put(5.8,.2){$\s 2$}
\put(6.8,.2){$\s 2$}
\put(8.8,.2){$\s 3$}
\put(.3,2.8){$\s 1$}
\put(.3,3.8){$\s 1$}
\put(4.3,1.8){$\s 2$}
\ep
\ea
$$
\\[1cm]
{\bf Acknowledgment:} The authors have profited from discussions with
A.~Fring, R.~Schra\-der, F.~Smirnov and A.~Belavin.

\end{document}